\documentclass[aps,prl,twocolumn,superscriptaddress,longbibliography,nofootinbib]{revtex4-2}
\usepackage{lmodern} 
\usepackage[utf8]{inputenc} 
\usepackage[T1]{fontenc} 
\usepackage{microtype} 
\usepackage[english]{babel}
\usepackage{color}
\usepackage{setspace}

\usepackage{mathbbol}

\usepackage{amsmath}
\usepackage{amssymb}
\usepackage{dsfont}
\usepackage{slashed}
\usepackage[usenames,dvipsnames]{xcolor}
\usepackage{changes}
\usepackage[colorlinks=true, linkcolor=black, citecolor=ForestGreen]{hyperref}

\newcommand{\ud}{\mathrm{d}}

\DeclareMathOperator\arctanh{arctanh}

\newcommand{\e}{\mathrm{e}} 
\newcommand{\h}{\text{h}}   
\begin{document}
\begin{flushright}
    CPHT-RR042.062025
\end{flushright}

\title{Near-extremal holographic charge correlators
}
\author{Blaise Gout\'eraux}
\email{blaise.gouteraux@polytechnique.edu}
\affiliation{CPHT, CNRS, \'Ecole polytechnique, IP Paris, F-91128 Palaiseau, France}
\author{David M.~Ramirez}
\email{dr634@cam.ac.uk}
\affiliation{CPHT, CNRS, \'Ecole polytechnique, IP Paris, F-91128 Palaiseau, France}
\affiliation{Department of Applied Mathematics and Theoretical Physics,
University of Cambridge, Cambridge CB3 0WA, UK}
\author{Mikel Sanchez-Garitaonandia}
\email{mikel.sanchez@polytechnique.edu}
\affiliation{CPHT, CNRS, \'Ecole polytechnique, IP Paris, F-91128 Palaiseau, France}
\author{Cl\'ement Supiot}
\email{clement.supiot@polytechnique.edu}
\affiliation{CPHT, CNRS, \'Ecole polytechnique, IP Paris, F-91128 Palaiseau, France}
\date{\today}

\begin{abstract}
We compute analytically the low-temperature charge correlators in near-extremal black holes with a planar horizon and an infrared AdS$_2\times \mathbb{R}^2$ extremal geometry, finding excellent agreement with numerical calculations. The analytical result consistently describes the crossover between the hydrodynamic diffusive regime at low frequencies and wavenumbers, and the quantum, zero-temperature regime at high frequencies and wavenumbers. We analytically resolve the successive collisions between the diffusive pole and the non-hydrodynamic poles sourced by the infrared AdS$_2\times \mathbb{R}^2$ geometry. We demonstrate that in the $T=0$ limit, a pair of gapless poles survive with a dispersion relation $\omega_\pm=-i k^2 + i (2/3) k^4(d_4\mp i\pi+\log k^2)$. The nonanalytic contributions arise from the interplay with the branch cut formed by the condensation of the non-hydrodynamic poles. The real part is caused by the `snatching' of one of the non-hydrodynamic poles by the hydrodynamic diffusive pole. 
\end{abstract}

\maketitle

\section{Introduction}

Holographic phases at finite density \cite{Zaanen:2015oix,Hartnoll:2016apf} capture numerous aspects of strongly-correlated quantum matter, such as superfluids \cite{Gubser:2008px,Hartnoll:2008vx,Hartnoll:2008kx}, Fermi surfaces \cite{Lee:2008xf,Liu:2009dm,Cubrovic:2009ye,Faulkner:2009wj} or metallic phases \cite{Andrade:2013gsa}. In the strong-coupling regime and in the absence of long-lived particles, the basic observables of interest are the many-body retarded Green's functions of conserved operators, and their calculation using holography has been the focus of many works over the years. 

The many-body retarded Green's functions give access to the spectrum of collective excitations, which characterizes the relaxation to equilibrium at various timescales. The spectrum can be computed numerically at any desired scale. However, analytical computations often yield more detailed insight into the underlying dynamics, emergent symmetries, etc. 
Analytical calculations of the spectrum usually rely on a small parameter expansion, such as in the hydrodynamic regime $\omega,k\ll T$. Barring special circumstances (self-duality \cite{Davison:2014lua}, pole skipping points \cite{Blake:2018leo,Arean:2020eus}, special limits \cite{Betzios:2017dol,Betzios:2018kwn,Gursoy:2021vpu}), the spectrum at $\omega\simeq T$ is out of analytical reach, as well as the crossover between the hydrodynamic regime and the `quantum' regime, $T\ll\omega,k$, and numerics must be employed.

In a recent series of works, poles of holographic retarded Green's functions have been determined through an approximate solution of the connection coefficient problem for the Heun equation, \cite{Bonelli:2022ten,Jia:2024zes,Ren:2024hwf,Arnaudo:2024sen}, to which the holographic equations governing them reduce in five bulk spacetime dimensions. Near extremality, \cite{Arnaudo:2024sen} observed that this approximation becomes exact in the zero-temperature limit, which allowed them to derive a formula for the correlator of a complex scalar operator and to access its pole structure. For a conserved current with integer scaling dimension, divergences arise. \cite{Jia:2024zes} argued that the correlator can still be extracted from the regular part of the expression obtained through this procedure.

In this work, we are specifically interested in the retarded Green's functions dual to perturbations around black hole spacetimes with a planar horizon, which in the extremal limit have an AdS$_2\times \mathbb{R}^2$ near-horizon geometry. In addition to playing an essential role in applications to holographic quantum matter, AdS$_2$ spacetimes have been established to be dual to Sachdev-Ye-Kitaev models of randomly interacting fermions \cite{Sachdev1993gapless,Georges2000,Sachdev:2010um,Kitaev2014hidden,Sachdev:2015efa,Maldacena:2016upp,Jensen:2016pah,Engelsoy:2016xyb} in the limit of low temperature and strong coupling. They are also prevalent in top-down String Theory constructions \cite{Ferrara:1995ih}.

The key result of this work is an analytical expression for the current-current retarded Green's function $G^\text{R}_{xx}(\omega,k)$ near extremality, in the regime $\omega\sim k^2\sim T\ll 1/r_\e$, where $r_\e$ is the extremal horizon radius and marks the limit of validity of the infrared effective theory:

\begin{equation}
    G^\text{R}_{xx} = 
    \frac{
        r_\e^{-1}\boldsymbol{\omega}^2\left(
            1 - \frac{\boldsymbol{k}^2\mathcal{G}}{3}
        \right)
    }{
        i \boldsymbol{\omega}
        + \frac{\boldsymbol{k}^2}6 \left( 
            4\pi \boldsymbol{T} 
            + 2(\boldsymbol{k}^2
            + i\boldsymbol{\omega})\mathcal G
            + 3\boldsymbol{k}^2\log 3
            - 6
        \right)
    },
 \label{green-function}
\end{equation}

\begin{equation}
  \mathcal{G}
  =\pi\cot\left(\frac{i\boldsymbol\omega}{2\boldsymbol T}\right)+\gamma+\psi\left(\frac{i\boldsymbol\omega}{2\pi\boldsymbol T}\right)-\log\left(\frac{9}{4\pi \boldsymbol T}\right),
  \label{mathcalG}
\end{equation}
with $\psi$ the digamma function, $\gamma$ the Euler's constant and where bold quantities are dimensionless in units of $r_\e$, for example $\boldsymbol \omega = \omega r_\e$ or $\boldsymbol{\zeta} = \zeta/r_\e$.

\begin{figure}
\begin{center}
    \includegraphics[width=0.5\textwidth]{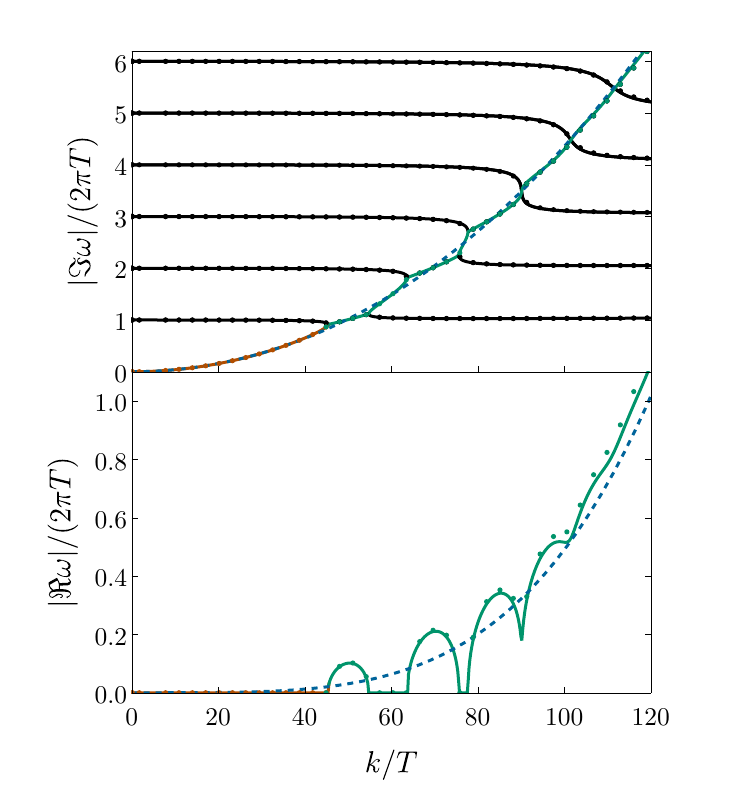}
\end{center}
\caption{\small (Top) Imaginary part and (Bottom) real part of the of the quasinormal modes for $T/m = 10^{-3}$. The poles of \eqref{green-function} are displayed as solid lines while the numerical result for the quasinormal modes are represented as dots. The gapless mode is colored in brown when in the hydrodynamic regime, while in green when out of it (after the collision with the first gapped mode). We have added the zero temperature analytical dispersion relation \eqref{zero-T-dispersion-relation} as a dashed blue line. }
      \label{fig:toverm-0.001}
\end{figure}

For $\omega,k\ll T$, the spectrum features a diffusive mode

\begin{equation}
    \frac{i\boldsymbol\omega}{\boldsymbol{k}^2} = \left(1-\frac{2\pi \boldsymbol T}3\right)  +\frac{\boldsymbol{k}^2}{6} \log\left[\frac{243}{16}\left(\frac{1}{2\pi \boldsymbol T}\right)^4\right]+\dots,
\end{equation}
where dots stand for higher $T$ and $k$ corrections. Moving beyond the hydrodynamic regime, \eqref{green-function} resolves the crossover to the quantum regime, and more specifically the collisions between the diffusive mode and a series of purely imaginary gapped poles. At small $k\ll r_\e^{-1}$, they are located at $\omega_n=-2 i \pi T(\Delta(k=0)+n)$, where $\Delta(\boldsymbol k)=1/2+1/2\sqrt{1+4\boldsymbol{k}^2/3}$ is the scaling dimension of the least irrelevant operator in the IR theory dual to AdS$_2\times \mathbb{R}^2$ \cite{Faulkner:2009wj}. They are a consequence of the SL$(2,\mathbb{R})$ emergent geometry of the AdS$_2$ black hole \cite{Faulkner:2011tm,Anninos:2011af}. A representative example is shown in figure \ref{fig:toverm-0.001}, demonstrating excellent agreement between the poles of \eqref{green-function} and numerical calculations. 

As $k/T$ increases, the charge diffusion pole undergoes a series of collisions with the infrared gapped poles, after which it moves out from the imaginary axis into the complex frequency plane, before recombining on the imaginary axis. This process repeats a few times until the quantum regime is reached, $T\ll k\ll r_\e^{-1}$, at which point the pair of poles remain complex. This is confirmed by taking the $T=0$ limit of \eqref{green-function}. The series of infrared poles condenses into a branch cut, while a pair of gapless poles survives with dispersion relation

\begin{equation}
    \boldsymbol\omega_\pm = 
    - i\boldsymbol{k}^2
    + i\frac23\boldsymbol{k}^4\left(
        d_4\mp i\pi+2\log (\boldsymbol k)
    \right),
\label{zero-T-dispersion-relation}
\end{equation}
where $d_4=\gamma+\log 2-\frac54\log3$.  Both the real $\mathcal O(k^4)$ and imaginary $\mathcal O(k^4\log k)$ terms can be traced back to the backreaction of the branch cut $(i\omega)^{2\Delta(k)}$ on the gapless mode: when expanded for small $k$, the branch cut ultimately gives $\log(i\omega)$ terms in the denominator of the retarded Green's function. The sign of the real part depends on the argument of $\omega$ compared to the logarithmic branch cut on the negative imaginary axis.

The appearance of a pair of gapless poles with a real part in the regime $T\ll k$ comes together with a downward shift of the levels of gapped infrared modes, as can be seen on the top panel of figure \ref{fig:toverm-0.001}. One of the infrared poles at $k\ll T$ gets `snatched' by the charge diffusion pole to form the poles \eqref{zero-T-dispersion-relation}, while maintaining consistency with constraints on Green's functions under parity transformations: poles are either purely imaginary or come with conjugate real parts \cite{Kaminski:2009dh}. This underlines the non-hydrodynamic nature of the gapless poles \eqref{zero-T-dispersion-relation}.

In the following, we describe the holographic model, the computation of the retarded Green's function and close with a discussion and outlook.

\section{Holographic model}

We study the $2+1$-dimensional decoupled dynamics of charge using a bulk action for a Maxwell gauge field,
\begin{equation}
\label{probeaction}
S=-\int \ud^4x\sqrt{-g}\,\frac{1}{4} F^2\,,
\end{equation}
with $F_{\mu\nu}=2\nabla_{[\mu}A_{\nu]}$ and indices $\mu, \nu=0,\dots 3$ being raised and lowered with the bulk metric $g_{\mu\nu}$. The metric is not dynamical, and we choose it to have UV AdS$_4$ asymptotics and an AdS$_2\times \mathbb{R}^2$ limit.

While we have obtained similar results hold for other spacetimes with an AdS$_2\times \mathbb{R}^2$ extremal geometry (such as Reissner-Nordstr\"om \cite{Faulkner:2009wj} or with a scalar deformation \cite{Blake:2016jnn}), we focus for analytical simplicity on a spacetime dual to a zero-density, momentum-relaxing, strongly-coupled metallic phase \cite{Bardoux:2012aw,Andrade:2013gsa,Davison:2014lua} and defer other cases to upcoming work \cite{Gouteraux:20252}. The model in the bulk is specified by adding $d=2$ massless scalar degrees of freedom,
\begin{equation}
\label{NeutralAction}
S_{\text{background}}=\int \ud^4x\sqrt{-g}\left[R-2\Lambda-\frac{1}{2}\sum_{i=1}^{2}\left(\partial\psi_i\right)^2\right], 
\end{equation}
where we have chosen units of in which $16\pi G_\text{N} = 1$. The solution to the classical equations of motion of this theory is homogeneous and known in closed form:
\begin{equation}
\begin{aligned}
    \ud s^2 & =\frac{\ell^2}{r^2}\left(-f(r)\ud t^2+\frac{\ud r^2}{f(r)}+\ud \vec x^2\right), \\
    f(r) & = 1-\frac{m^2r^2}{2}-\left(1-\frac{m^2 r_\h^2}{2}\right)\frac{r^3}{r_\h^3}, \quad \psi_i = m x^i,
\end{aligned}\label{axion-metric}
\end{equation}
with $\ell^2=-3/\Lambda$ the AdS$_4$ radius, $m$ the scale of momentum relaxation in the dual boundary theory. We will further choose units in which $\ell = 1$. The boundary is located at $r=0$, while the outer event horizon is at $r = r_\h$, which verifies $f(r_\h)=0$ and has temperature
\begin{equation}
    T = -\frac{f'(r_\h)}{4\pi} = \frac{6-m^2r_\h^2}{8\pi r_\h}.
    \label{temperature-axion}
\end{equation}
It vanishes in the extremal limit when the event horizon matches the inner horizon, $r_\h = r_\e =\sqrt{6}/m$.

The $T=0$ geometry corresponds to a domain wall connecting the UV AdS$_4$ and the IR AdS$_2 \times \mathbb{R}^2$, where the latter becomes manifest by changing to IR  coordinates that zoom into the near extremal horizon region \cite{Faulkner:2009wj},
\begin{equation}
    r = r_\e - \epsilon \frac{r_\e^2}{3\zeta},\quad r_\h = r_\e - \epsilon \frac{r_\e^2}{3\zeta_\h}, \quad t =\epsilon^{-1}\tau,
    \label{r-to-zeta}
\end{equation}
with $0<\zeta<\zeta_\h$ the radial coordinate of the inner region, in terms of which the extremal horizon is located at $\zeta,\zeta_\h \rightarrow\infty$. 
To leading order in $\epsilon$, the metric \eqref{axion-metric} becomes,
\begin{equation}
    \ud s^2 = 
    \frac{\ell_2^2}{\zeta^2}\left(
        - \left( 1 - \frac{\zeta^2}{\zeta_\h^2}\right)\ud\tau^2
        + \frac{\ud\zeta^2}{ 1 - \frac{\zeta^2}{\zeta_\h^2}}
    \right)
    + \frac{\ud \vec x^2}{r_\e^2},
    \label{metric-AdS2-finite-T}
\end{equation}
corresponding to an AdS$_2 \times \mathbb{R}^2$ black hole with radius given by $\ell_2 = 1/\sqrt{3}$. The boundary is formally located at $\zeta = 0$, but it lies beyond the range of validity of the IR coordinates $\zeta/r_\e \gtrsim\epsilon$ and a matching with the outer, AdS$_4$ region is required for smaller values of $\zeta$. The IR temperature is now given by $
T = \epsilon/(2\pi\zeta_\h)+\mathcal O(\epsilon^2)$.
The zero temperature AdS$_2\times \mathbb{R}^2$ geometry can be recovered by taking the $\zeta_\h\rightarrow\infty$ limit, and $\zeta_\h/r_\e \gtrsim\epsilon$ provides an upper bound on the highest temperatures which can be reached with \eqref{metric-AdS2-finite-T}. 

The current-current retarded Green's function is computed by solving for the perturbations of the bulk gauge field $\delta A_{\mu}(x^\nu)$. Since the background solution is homogeneous, we Fourier-decompose the perturbation in time and space,  $\delta A_{\mu} =\e^{-i(\omega t -k x)}a_{\mu}(r)$. Focusing on the longitudinal sector, the Maxwell equations decouple using the gauge-invariant electric field  $E_x = \omega a_x(r) +ka_t(r)$ \cite{Kovtun:2005ev}
\begin{equation}
    \left[\frac{f(r)E_x'}{\omega^2-k^2f(r)}\right]'+\frac{1}{f(r)}E_x=0.
    \label{longitudinal-Ex-equation}
\end{equation}
Once a solution is obtained, we extract the current-current retarded Green's function through
\begin{equation}
\label{GRxx}
    G^\text{R}_{xx}(\omega,k;T) = \frac{\omega^2}{\omega^2-k^2}\left.\frac{\partial_r E_x}{E_x}\right\vert_{r=0}.
\end{equation}
Other correlators such as the density-density $G^\text{R}_{tt}$ can be obtained using the charge Ward identity.

We solve this equation by matching a solution valid in the outer region defined by $r^2\omega^2/f^2,r^2k^2/f\ll1$ \cite{Davison:2013bxa}, to a solution valid in the inner region, $\zeta/r_\e,\zeta_\h/r_\e \gtrsim\epsilon$. This procedure can be streamlined by constructing both solutions order by order in $\epsilon$, $E_x^{(\mathcal{O})}(r) = \sum_{n\geq 0}\epsilon^{n} E^{(\mathcal{O})}_{x,n}(r)$ and $E_x^{(\mathcal{I})}(\zeta) = \sum_{n\geq 0}\epsilon^{n} E^{(\mathcal{I})}_{x,n}(\zeta)$. In fact, the conditions $r^2\omega^2/f^2,r^2k^2/f\ll1$ imply that in the outer region, we scale $\omega\sim k^2\sim\epsilon$, while in the inner region $\omega\sim k^2\sim T\sim\epsilon$. Temperature corrections do not appear in the outer equations up to order $\mathcal O(\epsilon^3)$, so to the order we are interested in here, we can simply construct the outer solutions setting the temperature to zero.

\section{Outer solution}

Formally, the solution to \eqref{longitudinal-Ex-equation} can be written as
\begin{equation}
\begin{split}
    E_x = &
    \mathcal{E}_0 
    + \frac{\mathcal{E}_1}{k^2 - \omega^2}\left(k^2 r - \omega^2\int_0^r\frac{\ud r_1}{f(r_1)} \right)\\&
    + \int_0^r \ud r_1\left( k^2 - \frac{\omega^2}{f(r_1)} \right)
    \int_0^{r_1}\ud r_2\frac{E_x(r_2)}{f(r_2)},
\end{split}
    \label{formal-outer-solution}
\end{equation}
with integration constants $\mathcal{E}_0$ and $\mathcal{E}_1$. By scaling $\mathcal{E}_0 = \sum_{n\geq 0}\epsilon^{n}\mathcal{E}_{0}^{(n)}$ and $\mathcal{E}_1 = \sum_{n\geq 0}\epsilon^{n}\mathcal{E}_{1}^{(n)}$, together with $\omega\sim k^2\sim\epsilon$, we obtain the outer solution iteratively order by order. To consistently obtain the dispersion relation to $\mathcal{O}(\epsilon^2)$, it is necessary to work out the outer solution to $\mathcal O(\epsilon^2)$ included. Fortunately, temperature corrections only show up in the outer equations at order $\mathcal O(\epsilon^3)$, which simplifies the analytical expressions\footnote{This fact holds for the planar Reissner-Nordström black hole too.}. To match the solutions, we expand them in the overlap region using \eqref{r-to-zeta}. This leads to mixing in the $\epsilon$ orders, i.e.~the $\mathcal{O}(\epsilon^{n>0})$ solution generates $\mathcal{O}(\epsilon^{n-1})$ terms.
At order $\mathcal O(\epsilon^0)$,
\begin{equation}
     E_{x,0}^{(\mathcal{O})}(r) = \mathcal{E}_{0}^{(0)}+\mathcal{E}_{1}^{(0)}r.
\end{equation}
Higher-order expressions are more complicated and we give them in the End Matter.

\section{Inner solution}

We obtain the inner solutions by scaling the equation \eqref{longitudinal-Ex-equation} as $\omega\sim k^2\sim \epsilon$ and changing coordinates to IR coordinates \eqref{r-to-zeta}. 

In what follows, we will work with the bold, dimensionless variables.
Order by order in $\epsilon$, we obtain a set of equations
\begin{equation}
\label{innereq}
     \mathcal{D}E^{(\mathcal{I})}_{x,n\geq0} 
     = \mathcal S^{(n)}[E^{(\mathcal{I})}_{x,n-1},\dots E^{(\mathcal{I})}_{x,0}],
\end{equation}
where the homogeneous differential operator 

\begin{equation}
      \mathcal{D}E^{(\mathcal{I})}_{x,n} \equiv 
     \partial^2_{\boldsymbol\zeta}E^{(\mathcal{I})}_{x,n}
     - \frac{2\boldsymbol\zeta}{\boldsymbol\zeta_\h^2}
     \frac{
        \partial_{\boldsymbol\zeta}E^{(\mathcal{I})}_{x,n}
    }{
        1-(\boldsymbol\zeta/\boldsymbol\zeta_\h)^2
    }
     + \frac{
        \boldsymbol{\omega}^2E^{(\mathcal{I})}_{x,n}
     }{
        (1-(\boldsymbol\zeta/\boldsymbol\zeta_\h)^2)^2
    }
\end{equation}
and the inhomogeneous source $\mathcal S^{(0)}=0$. We give some explicit expressions in the End Matter. 

The solution to the zeroth-order inner equation \eqref{innereq} is
\begin{equation}
    E^{(\mathcal{I})}_{x,0} = c_\pm^{(0)}\e^{\pm i\boldsymbol\omega\boldsymbol\zeta_\h\arctanh\xi},
\end{equation}
with $c_{\pm}^{(0)}$ integration constants and $\xi=\boldsymbol\zeta/\boldsymbol\zeta_\h$. We now impose ingoing boundary conditions on this solution \cite{Son:2002sd}. This is done by changing the radial coordinate to the tortoise coordinate $r^\star(r)=\int_0^r \ud r_1/f(r_1)$, \eqref{tortoiseintegral} in the End Matter, and then to ingoing Eddington-Finkelstein coordinate $t\mapsto v+r^\star$ so that $\delta A_\mu\sim \e^{-i\omega v}$ near the horizon. To leading order, this imposes $c_{-}^{(0)}=0$. This procedure can be repeated to $\mathcal O(\epsilon)$, with the additional subtlety that the tortoise coordinate \eqref{tortoiseintegral} must also be expanded in $\epsilon$:

\begin{equation}
\begin{aligned}
    E_x^{(\mathcal{I})}\underset{\boldsymbol\zeta\to\boldsymbol\zeta_\h}{=}&
    E^{(\mathcal{I})}_{x,0}\Big\{
        1 + \epsilon\frac{i\boldsymbol{\omega}}{36}\Big[ 
            3\log(1-\xi)-2\frac{6-\xi}{1+\xi}\\
            & 
            + 8\log (9\boldsymbol\zeta)
            - 11\log(1+\xi)
        \Big]
        +\mathcal{O}(\epsilon^2)
    \Big\}.
\end{aligned}\label{infalling-small-T}
\end{equation}
We present the expression for the first-order inner solution in the End Matter.

\section{Matching and Green's function}

With both the outer and inner solutions in hand, we can now match them. This is done by consistently expanding to the same order in $\epsilon$ both $E_x^{\mathcal{(O)}}$  after changing to infrared coordinates \eqref{r-to-zeta}, and $E_x^{\mathcal{(I)}}$ after expanding in $\zeta\to0$. The outcome of this procedure is that $\mathcal E_0$ and $\mathcal E_1$, the integration constants of the outer solution, become fixed in terms of the inner integration constant $c_+^{(0)}$:

\begin{equation}
    \mathcal{E}_0^{(0)} = 
    c_+^{(0)} \left(
        1
        - \frac{\boldsymbol{k}^2}{i\boldsymbol\omega}
    \right),
    \quad 
    \mathcal{E}_1^{(0)} = 
    c_+^{(0)} r_\e^{-1} \frac{\boldsymbol{k}^2}{i\boldsymbol\omega},
    \label{2nd-order-matching}
\end{equation}
\begin{equation}
    \begin{aligned}
        \mathcal{E}_0^{(1)} = &~
        \frac{c_+^{(0)}\boldsymbol{k}^2}{6i\boldsymbol\omega\boldsymbol\zeta_\h}
        \left[
            2 + 
            3\boldsymbol k^2\boldsymbol\zeta_\h\log 3
            + 2\boldsymbol\zeta_\h\left(\boldsymbol{k}^2
            + i\boldsymbol\omega\right)\mathcal{G}
        \right],\\
        \mathcal{E}_1^{(1)} = &~
        c_+^{(0)}i\boldsymbol\omega \,r_\e^{-1}\left(
            1 + \frac{\boldsymbol{k}^4\mathcal{G}}{3\boldsymbol{\omega}^2}
        \right),
    \end{aligned}\label{1st-order-matching}
\end{equation}
where $\mathcal G$ is given in \eqref{mathcalG}. The Green's function \eqref{green-function} is obtained by inserting the outer solution in \eqref{GRxx}, taking the near-boundary limit and using $\zeta_\h=\epsilon/(2\pi T)+\mathcal O(\epsilon^2)$.

\section{Discussion and outlook}

In this work, we have focused on the simple analytical solution \eqref{axion-metric}. In upcoming work \cite{Gouteraux:20252}, we extend this analysis to other cases with AdS$_2\times \mathbb{R}^2$ near-horizon extremal geometries, such as the Reissner-Nordstr\"om solution or black holes with scalar deformations \cite{Blake:2016jnn}, for which we have found qualitatively similar results. Our results can essentially be restated as the fact that, in the small-temperature limit, we are able to write down an effective theory which captures both gapless and gapped modes. Here we have only computed the retarded Green's functions of this theory. However, the effective action of the theory can be computed using the approach of \cite{Nickel:2010pr,Davison:2022vqh}
 (see also \cite{Liu:2024tqe}). This would reveal any emergent symmetries and also clarify the origin of the zero-temperature gapless poles \eqref{zero-T-dispersion-relation}. Extending this computation to the complexified gravitational Schwinger-Keldysh contour constitutes another step towards fully determining the effective action of these holographic systems and would pave the way towards capturing their stochastic properties, \cite{Glorioso:2018mmw, Jana:2020vyx,Ghosh:2020lel,He:2021jna,He:2022deg,He:2022jnc,Loganayagam:2022zmq}.

To place this into context, so far hydrodynamic theories with gapped poles have been worked out when these poles are caused by the weak breaking of a global symmetry; see e.g. \cite{Davison:2014lua,Grozdanov:2018fic,Baggioli:2022pyb,Gouteraux:2024adm}. For instance, if spatial translations are weakly broken and momentum conservation is approximate, then one is able to write down an effective theory perturbative in the strength of the coupling breaking translations. Accordingly, this theory continuously reduces to hydrodynamics with conserved momentum when the coupling is turned off. The gapped pole location is proportional to the coupling, and the pole becomes gapless in the limit of zero coupling. Here, the situation is different: the gapped poles are proportional to temperature, and the small coupling is temperature itself.

Our results can be extended to include backreaction. There, zero-temperature gapless poles are also known to arise for the Reissner-Nordstr\"om black brane \cite{Edalati:2010hk,Edalati:2010pn,Davison:2013bxa,Moitra:2020dal}, in the spacetime \eqref{axion-metric} \cite{Arean:2020eus} and in spacetimes with a near-extremal geometry conformal to AdS$_2\times \mathbb{R}^2$ \cite{Davison:2013uha,Liu:2021qmt}. Computing the low-temperature Green's function and even the effective action describing these modes would be a significant step towards elucidating their origin. The fluid/gravity treatment in \cite{Moitra:2020dal} suggests that this is within reach. Interestingly, both in the Reissner-Nordstr\"om black brane and in the spacetime \eqref{axion-metric} \cite{Arean:2020eus}, the pole collisions only occur for imaginary wavenumber and non-analytic contributions to the dispersion relation of the gapless poles appear to be suppressed. This is in contrast to the results presented here or in the SYK-chain model \cite{Choi:2020tdj}. 

We expect that the energy sector in the backreacted case is controlled by the Schwarzian action, when the near-extremal geometry has an emergent AdS$_2$ factor \cite{Kitaev2014hidden,Maldacena:2016hyu,Maldacena:2016upp,Jensen:2016pah,Engelsoy:2016xyb}. Such gapless modes also appear in other sectors, like the shear sector in Reissner-Nordstr\"om \cite{Edalati:2010hk,Edalati:2010pn,Davison:2013bxa}). The role played by the Schwarzian action and its interplay with higher energy degrees of freedom remains to be elucidated.

Finally, the question of whether such zero-temperature gapless modes survive quantum corrections should be addressed. This can be tackled by coupling their effective action to the full quantum Schwarzian action,  \cite{Iliesiu:2020qvm,Iliesiu:2022onk}. Recently, quantum corrections to the absorption cross-section of gravitons have been computed, \cite{Emparan:2025sao}, as well as to the two-point function of a probe fermion, \cite{Liu:2024gxr}.
\newline

\begin{acknowledgments}
We are grateful to Richard Davison, Roberto Emparan, Elias Kiritsis and Eric Mefford for discussions, and to Richard Davison for comments on the manuscript. B.~G.~ gratefully acknowledges support from the Simons Center for Geometry and Physics, Stony Brook University, and the organizers of the program `Black hole physics from strongly coupled thermal dynamics' at which this paper was finalized. B.~G., D.~M.~R.~ and M.~S-G. were partially supported during this work by the European Research Council (ERC) under the European Union's Horizon 2020 research and innovation programme (grant agreement No758759). The work of D.~M.~R.~is partially supported by STFC consolidated grant ST/X000664/1 and by the Simons Investigator award $\#$620869.
\end{acknowledgments}

\bibliography{biblioAdS2}
\onecolumngrid
\appendix
\section{End Matter}
\subsection{Outer solution}

The outer solution is
\begin{equation}
    E_x^{(\mathcal{O})}(r) = \sum_{n\geq 0}\epsilon^{n} E^{(\mathcal{O})}_{x,n}(r),
\end{equation}
where the solutions at each order are
$   E_{x,0}^{(\mathcal{O})}(r) = \mathcal{E}_{0}^{(0)}+\mathcal{E}_{1}^{(0)}r,$
and 
\begin{equation}
    E_{x,1}^{(\mathcal{O})}(r) = \mathcal{E}_{0}^{(1)} + \left(\mathcal{E}_{1}^{(1)}+ \frac{\omega^2}{k^2}\mathcal{E}_{1}^{(0)}\right) r  - \frac{\omega^2}{k^2}\mathcal{E}_{1}^{(0)} \int_0^r\frac{\ud r_1}{f(r_1)}+ k^2\int_0^r\ud r_1\int_0^{r_1}\ud r_2\frac{E_{x,0}^{(\mathcal{O})}(r_2)}{f(r_2)},
\label{sol-hydro-1st}
\end{equation}

\begin{equation}
\begin{aligned}
    E_{x,2}^{(\mathcal{O})}(r) = &~  \mathcal{E}_{0}^{(2)}+\left[\mathcal{E}_{1}^{(2)}+\frac{\omega^2}{k^2}\left(\mathcal{E}_1^{(1)}+\frac{\omega^2}{k^2}\mathcal{E}_1^{(0)}\right)\right]r - \frac{\omega^2}{k^2}\left(\mathcal{E}_1^{(1)}+\frac{\omega^2}{k^2}\mathcal{E}_1^{(0)}\right)\int_0^r \frac{\ud r_1}{f(r_1)}\\
    &~ -\omega^2\int_0^r\frac{\ud r_1}{f(r_1)}\int_0^{r_1}\ud r_2\frac{E_{x,0}^{(\mathcal{O})}(r_2)}{f(r_2)} + k^2\int_0^r\ud r_1\int_0^{r_1}\ud r_2\frac{E_{x,1}^{(\mathcal{O})}(r_2)}{f(r_2)},
\end{aligned}\label{sol-hydro-2nd}
\end{equation}

Some of these integrals are required to match the inner with the outer solution. 

As we did in the inner region, we will represent in bold those variables normalized by $r_\e$. 
Setting $\boldsymbol r_\h = 1$ and $\boldsymbol m = \sqrt{6}$, we arrive to the following expressions:

\begin{equation}
\label{tortoiseintegral}
  \int_0^{\boldsymbol r}\frac{\ud \boldsymbol r_1}{f} = \frac{\boldsymbol r}{3(1-\boldsymbol r)}+\frac{4}{9}\arctanh\left(\frac{3\boldsymbol r}{2+\boldsymbol r}\right),\quad   \int_0^{\boldsymbol r}\ud \boldsymbol r_1\int_0^{\boldsymbol r_1}\frac{\ud \boldsymbol r_2}{f} = -\frac{\boldsymbol r}{3}+\frac{2(1+2\boldsymbol r)}{9} \arctanh\left(\frac{3\boldsymbol r}{2+\boldsymbol r}\right),
\end{equation}
\begin{equation}
    \begin{split}
    \int_0^{\boldsymbol r}\ud \boldsymbol r_1\int_0^{\boldsymbol r_1}\ud \boldsymbol r_2\frac{\boldsymbol r_2}{f} =& \frac{1}{18}\left[-6r-\log(2\boldsymbol r+1)-4\boldsymbol r\arctanh\left(\frac{3\boldsymbol r}{2+\boldsymbol r}\right)-8\log(1-\boldsymbol r)\right],
    \end{split}
\end{equation}
\begin{equation}
    \begin{split}
        \int_0^{\boldsymbol r} \frac{\ud \boldsymbol r_1}{f}
        \int_0^{\boldsymbol r_1}\frac{\ud \boldsymbol r_2}{f} =& 
        \frac{
            \left(
                3\boldsymbol r
                + 4(1-\boldsymbol r)\arctanh\left(
                    \frac{3 \boldsymbol r}{2 + \boldsymbol r}
                \right)
            \right)^2
        }{
            162(1-\boldsymbol r)^2
        },
    \end{split}
\end{equation}
\begin{equation}
    \begin{split}
    \int_0^{\boldsymbol r}\frac{\ud \boldsymbol r_1}{f}\int_0^{\boldsymbol r_1}\ud \boldsymbol r_2\frac{\boldsymbol r_2}{f} =& \frac{1}{162}\left[-9+\frac{9}{(1-\boldsymbol r)^2} -\frac{12\arctanh\left(\frac{3 \boldsymbol r}{2+\boldsymbol r}\right)}{1-\boldsymbol r} -8\arctanh\left(\frac{3 \boldsymbol r}{2+\boldsymbol r}\right)^2\right],
    \end{split}
\end{equation}
\begin{equation}
    \begin{split}
    \int_0^{\boldsymbol r}\ud \boldsymbol r_1\int_0^{\boldsymbol r_1}\frac{\ud \boldsymbol r_2}{f}\int_0^{\boldsymbol r_3}\frac{\ud \boldsymbol r_3}{f} = &\frac{1}{162}\left\{\frac{9\boldsymbol r(2-\boldsymbol r)}{1-\boldsymbol r} +4(1+2\boldsymbol r)\arctanh\left(\frac{3 \boldsymbol r}{2+\boldsymbol r}\right)\left[2\arctanh\left(\frac{3 \boldsymbol r}{2+\boldsymbol r}\right)-3\right]\right\}.
    \end{split}
\end{equation}

The expansion of the outer solutions in the matching region is

\begin{equation}
\begin{aligned}
        E^{(\mathcal{O})}_{x,0} =&~ 
        \mathcal{E}_0^{(0)} 
        + \mathcal{E}_1^{(0)}r_\e
        - \epsilon \frac{\mathcal{E}_1^{(0)}r_\e}{3\boldsymbol\zeta} 
        + \mathcal{O}(\epsilon^2),\\
        \\
        E^{(\mathcal{O})}_{x,1} =&~ 
        - \epsilon^{-1}
        \frac{\boldsymbol\omega^2}{\boldsymbol k^2}
        \mathcal{E}_1^{(0)}r_\e
        \boldsymbol\zeta
        + \mathcal{E}_0^{(1)}
        + \mathcal{E}_1^{(1)}r_\e
        +\frac{2\boldsymbol\omega^2}{9 \boldsymbol k^2}\left(
            6-2\log 3
        \right) \mathcal{E}_1^{(0)}r_\e
        + \frac{\boldsymbol k^2}{6}\left[
            2(2\log 3 - 1) \mathcal{E}_0^{(0)}
            + (\log 3 - 2) \mathcal{E}_1^{(0)}r_\e
        \right]\\
        & + \left[
            \frac{\boldsymbol k^2}{3}\left(
                \mathcal{E}_0^{(0)} + \mathcal{E}_1^{(0)}r_\e
            \right)
            - \frac{2\boldsymbol\omega^2}{9 \boldsymbol k^2}\mathcal{E}_1^{(0)} r_\e
        \right]\log\boldsymbol\zeta
        + \mathcal{O}(\epsilon),\\
        \\
        E^{(\mathcal{O})}_{x,2} =&~ 
        - \frac{\epsilon^{-2}}{2}\left(
            \mathcal{E}_0^{(0)}
            + \mathcal{E}_1^{(0)} r_\e
        \right)\boldsymbol\omega^2\boldsymbol\zeta^2
        - \frac{\boldsymbol\zeta}{\epsilon} \left\{
            \frac{\boldsymbol\omega^4}{\boldsymbol k^4} \mathcal{E}_1^{(0)} r_\e
            + \frac{\boldsymbol\omega^2\mathcal{E}_1^{(1)}r_\e}{\boldsymbol k^2}
            + \frac{\boldsymbol\omega^2}{18}\left(4\log 3 - 3\right) \left(
                2\mathcal{E}_0^{(0)}
                - \mathcal{E}_1^{(0)}r_\e
            \right)
        \right.\\
        &~\left. 
            + \frac{\boldsymbol\omega^2}{9}\left(
                2\mathcal{E}_0^{(0)}
                - \mathcal{E}_1^{(0)}r_\e
            \right) \log\boldsymbol\zeta
        \right\}
        + \mathcal{O}(\epsilon^0).
\end{aligned}\label{expansion-outer}
\end{equation}
Notice here that both $E^{(\mathcal{O})}_{x,1}$ and $E^{(\mathcal{O})}_{x,2}$ give terms of order $\mathcal O(\epsilon^{-1})$ and $\mathcal O(\epsilon^{0})$, while $E^{(\mathcal{O})}_{x,2}$ also contributes a term of order $\mathcal O(\epsilon^{-2})$.

\subsection{Inner solution}
The inhomogeneous source for the first order inner solution is (defining $\xi=\boldsymbol\zeta/\boldsymbol\zeta_\h$ and remembering that bold characters refer to variables normalized with $r_\e$, 

\begin{equation}
    \begin{aligned}
        \mathcal{S}^{(1)}[E^{(\mathcal{I})}_{x,0}] =&
        -\frac{2}{9\boldsymbol\zeta^2}\left(
            1 + \frac{7}{2}\frac{\xi^2}{(1+\xi)^2} 
            - \frac{3 \boldsymbol k^2}{\boldsymbol\omega^2\boldsymbol\zeta}
        \right) \partial_{\boldsymbol\zeta}E^{(\mathcal{I})}_{x,0}
        - \left(
            \frac{
                4\boldsymbol\omega^2\left( 1 + \xi - 7\xi^2/2 \right)
            }{
                9\boldsymbol\zeta (1+\xi) \left( 1 - \xi^2 \right)^2
            }
            -\frac{\boldsymbol k^2}{3\boldsymbol\zeta^2}\frac{1}{1-\xi^2}
        \right) E^{(\mathcal{I})}_{x,0}.
    \end{aligned}
\end{equation}
The corresponding first order solution is then

\begin{equation}
    \begin{aligned}
        E^{(\mathcal{I})}_{x,1} = & 
        c_+^{(1)}\e^{i\boldsymbol\omega\boldsymbol\zeta_\h\arctanh\xi} 
        + c_-^{(1)}\e^{-i\boldsymbol\omega\boldsymbol\zeta_\h\arctanh\xi} + c_+^{(0)}\e^{i\boldsymbol\omega\boldsymbol\zeta_\h\arctanh\xi}\left[
            \frac{\boldsymbol k^2}{3i\boldsymbol\omega\boldsymbol\zeta}\left(
                \frac{1}{2i\boldsymbol\omega\boldsymbol\zeta}
                -1
            \right)
        \right.\\
        &\left.  + \frac{7}{18}\left(
                \frac{1}{\boldsymbol\zeta_\h}
                - \frac{i\boldsymbol{\omega}}{1+\xi}
            \right)
            + \frac{i\boldsymbol{\omega}}{36}\left(
                3\log\frac{1-\xi}{\xi}
                - 11\log\frac{1+\xi}{\xi}
            \right)
        \right.\\
        &\left.
            - \frac{\boldsymbol{k}^2}{6}\left(
                \frac{_2F_1\left[
                    1,i\boldsymbol\omega\boldsymbol\zeta_\h;1+i\boldsymbol\omega\boldsymbol\zeta_\h;\frac{1+\xi}{1-\xi}
                \right]}{i\boldsymbol\omega\boldsymbol\zeta_\h}
                 + \frac{1+\xi}{1-\xi} 
                \frac{_2F_1\left[
                    1,1+i\boldsymbol\omega\boldsymbol\zeta_\h;2+i\boldsymbol\omega\boldsymbol\zeta_\h;\frac{1+\xi}{1-\xi}
                \right]}{1+i\boldsymbol\omega\boldsymbol\zeta_\h}
            \right)
        \right],
    \end{aligned}
\end{equation}
with $c_{\pm}^{(1)}$ being the new integration constants of the homogeneous solution and the remaining terms correspond to the particular solution with source turned on by $E_{x,0}$. In order to fix the new integration constants, we must carefully take care of the different terms as the $\xi\rightarrow 1$ limit is taken. 

Let us start by taking a look into the hypergeometric functions. Near the horizon, their combination behaves as,

\begin{equation}
    \begin{aligned} 
    & \e^{i\boldsymbol\omega\boldsymbol\zeta_\h \arctanh\xi} \left(
        \frac{
            _2F_1\left[
                1,
                i\boldsymbol\omega\boldsymbol\zeta_\h;
                1+i\boldsymbol\omega\boldsymbol\zeta_\h;
                \frac{1+\xi}{1-\xi}
            \right]}
        {
            i\boldsymbol{\omega\zeta}_\h
        }
        + \frac{1+\xi}{1-\xi}
        \frac{
            \,_2F_1\left[
                1,
                1+i\boldsymbol\omega\boldsymbol\zeta_\h;
                2+i\boldsymbol\omega\boldsymbol\zeta_\h;
                \frac{1+\xi}{1-\xi}
            \right]
        }{
            1 + i\boldsymbol{\omega\zeta}_\h
        }
    \right)\\
    & = 2^{-i\boldsymbol\omega\boldsymbol\zeta_\h/2}\left[
        2\pi\bigg(
            -i + \cot(i\pi\boldsymbol\omega\boldsymbol\zeta_\h)
            +\mathcal{O}(1-\xi)
        \bigg) \left(1-\xi\right)^{i\boldsymbol\omega\boldsymbol\zeta_\h/2}
        - \left(
            \frac{
                2^{i\boldsymbol\omega\boldsymbol\zeta_\h}
            }{
                i\boldsymbol\omega\boldsymbol\zeta_\h
            }+\mathcal{O}(1-\xi)
        \right) \left(1-\xi\right)^{-i\boldsymbol\omega\boldsymbol\zeta_\h/2}
    \right],
    \end{aligned}
\end{equation}
which is given by a combination of both ingoing and outgoing modes. In order to get rid of the latter, we must impose 

\begin{equation}
    c_-^{(1)} = \frac{c_+^{(0)}}{3}\pi \boldsymbol k^2\left(-i+\cot(i\pi\boldsymbol\omega\boldsymbol\zeta_\h)\right).
\end{equation}
On the other hand, $c_+^{(1)}$ has to be tuned in order to give the precise terms that arise at next to leading order in \eqref{infalling-small-T}:

\begin{equation}
   c_+^{(1)} = c_+^{(0)}\left[-\frac{7}{18\boldsymbol\zeta_\h}+\frac{\boldsymbol k^2}{6i\boldsymbol\omega\boldsymbol\zeta_\h}\left(1-\frac{1}{i\boldsymbol\omega\boldsymbol\zeta_\h}\right)+\frac{i\boldsymbol\omega}{18}\left(1+8\log 3+4\log\boldsymbol\zeta_\h\right)\right].
\end{equation}

Once the integration constants have been fixed, we can perform the $\zeta\rightarrow 0$ expansion required to perform the matching between the outer and inner region solutions. The expansions are,

\begin{equation}
\begin{aligned}
    E^{(\mathcal{I})}_{x,0} = & c_+^{(0)}\left(1+i\boldsymbol\omega\boldsymbol\zeta-\frac{\boldsymbol\omega^2}{2}\boldsymbol\zeta^2+\mathcal{O}(\boldsymbol{\zeta}^3)\right),
    \\
    E^{(\mathcal{I})}_{x,1} = &~ 
    c_+^{(0)} \left\{
        \frac{i \boldsymbol k^2}{3\boldsymbol\omega}\boldsymbol\zeta^{-1}
        + \left(\frac{\boldsymbol k^2}{3}
        + \frac{2i\boldsymbol\omega}{9}\right)\log\boldsymbol\zeta
        - i\boldsymbol\omega\left(
            \frac{1}{3}
            - \frac{4}{9}\log 3
        \right)
        +\frac{\boldsymbol k^2}{3}\left(
            - 1
            +\frac{1}{i\boldsymbol\omega\boldsymbol\zeta_\h}
            +2\log 3 + \mathcal{G}(\boldsymbol\zeta_\h,\boldsymbol\omega) 
        \right)
    \right.\\
    &~ \left. \qquad +
        \boldsymbol\zeta\left[
            - \left( 
                \frac{i\boldsymbol\omega k^2}{3}
                + \frac{2\boldsymbol\omega^2}{9}
            \right) \log{\boldsymbol\zeta}
            + \boldsymbol\omega^2\left(
                \frac{1}{3}
                - \frac{4}{9}\log 3
            \right)
            -\frac{i\boldsymbol\omega k^2}{3}\left(
                -\frac{3}{2}
                +2\log 3 + \mathcal{G}(\boldsymbol\zeta_\h,\boldsymbol\omega)
            \right)
        \right]
    \right\} + \mathcal{O}\left(\boldsymbol\zeta^2\right).
\end{aligned}\label{expansion-inner}
\end{equation}

\end{document}